\documentclass[journal]{IEEEtran}

\usepackage{amssymb}
\usepackage[cmex10]{amsmath}
\usepackage{cite}
\usepackage{graphicx,epsfig}
\usepackage{xcolor,colortbl}
\usepackage{booktabs}
\usepackage{makecell}
\usepackage{array,multirow}
\newcolumntype{P}[1]{>{\centering\arraybackslash}p{#1}}

\usepackage{hhline}
\usepackage{multirow}
\usepackage{hhline}
\usepackage{amsmath}
\usepackage{bbm}
\usepackage{epstopdf}
\usepackage{color}

\newcommand{\ua}{\uparrow}

\newcommand{\nc}{\newcommand}
\nc{\da}{\downarrow} \nc{\hc}{\hat{c}} \nc{\hS}{\hat{S}}
\nc{\bra}{\langle} \nc{\ket}{\rangle} \nc{\eq}{equation (\ref}
\nc{\h}{\hat} \nc{\hT}{\h{T}}\nc{\be}{\begin{eqnarray}}
\nc{\ee}{\end{eqnarray}}\nc{\rd}{\textrm{d}}\nc{\e}{eqnarray}\nc{\hR}{\hat{R}}\nc{\Tr}{\mathrm{Tr}}
\nc{\tS}{\tilde{S}}\nc{\tr}{\mathrm{tr}}\nc{\8}{\infty}\nc{\lgs}{\bra\ua,\phi|}\nc{\rgs}{|\ua,\phi\ket}
\nc{\hU}{\hat{U}}\nc{\lfs}{\bra\phi|}\nc{\rfs}{|\phi\ket}\nc{\hZ}{\hat{Z}}\nc{\hd}{\hat{d}}\nc{\mD}{\mathcal{D}}
\nc{\bd}{\bar{d}}\nc{\bc}{\bar{c}}\nc{\mc}{\mathcal}\nc{\ea}{eqnarray}\nc{\mG}{\mathcal{G}}\nc{\bce}{\begin{center}}
\nc{\ece}{\end{center}}
\nc{\heffl}{\pmb{H}_{\text{eff}}}
\nc{\lap}{\text{Laplacian}}
\nc{\iso}{\text{isotropic}}
\nc{\dlap}{d_\lap}
\nc{\diso}{d_\iso}
\nc{\ptp}{p_\upsilon}
\nc{\rtp}{{\pmb{R}^{1/2}_n}}
\nc{\thh}{\text{th}}

\date{September 2014}
\begin{document}

\title{Relationship between Cross-Polarization Discrimination (XPD) and Spatial Correlation in Indoor Small-Cell MIMO Systems}

\author{Yeon-Geun Lim,~\IEEEmembership{Student~Member,~IEEE,} Yae Jee Cho,~\IEEEmembership{Student~Member,~IEEE,} TaeckKeun Oh,~\IEEEmembership{Member,~IEEE,} Yongshik Lee,~\IEEEmembership{Senior~Member,~IEEE}, and Chan-Byoung Chae,~\IEEEmembership{Senior~Member,~IEEE}
\thanks{Y.-G. Lim, Y. J. Cho, and C.-B. Chae are with School of Integrated Technology, Yonsei University, Korea (e-mail: {yglim, yjenncho, cbchae}@yonsei.ac.kr). T. Oh is with LIG Nex1, Korea (e-mail: taeckkeun.oh@lignex1.com). Y. Lee is with School of Electrical and Electronic Engineering, Yonsei University, Korea (e-mail: yongshik.lee@yonsei.ac.kr).}
\thanks{This research was supported by the MSIP (Ministry of Science, ICT and Future Planning), Korea, under the ``IT Consilience Creative Program'' (IITP-R0346-16-1008) supervised by the IITP and ICT R\&D program of MSIP/IITP (2015-0-00300).}
}

\maketitle

\begin{abstract}
	
In this letter, we present a correlated channel model for a dual-polarization antenna to omnidirectional antennas in indoor small-cell multiple-input multiple-output (MIMO) systems. In an indoor environment, we confirm that the cross-polarization discrimination (XPD) in the direction of angle-of-departure can be represented as the spatial correlation of the MIMO channel. We also evaluate a dual-polarization antenna-based MIMO channel model and a spatially correlated channel model using a three-dimensional (3D) ray-tracing simulator. Furthermore, we provide the equivalent distance between adjacent antennas according to the XPD, providing insights into designing a dual-polarization antenna and its arrays.\\
\end{abstract}


\section{Introduction}\label{Sec:Intro}
Researchers have been able to extend mobile service coverage and network capacity through their development of small-cell technology \cite{Jang_WCM16}. In next-generation communications, researchers are considering multiple-input multiple-output (MIMO) techniques\cite{Chae_SPMag_07}. One such technique is massive MIMO, in which base stations (BSs) are equipped with many antennas to increase capacity and to conserve energy \cite{Rusek_SPMag_12, Lim_TWC}. To establish MIMO schemes in small-cell systems, researchers must address certain issues. MIMO capacity can be degraded, for example, by a compact antenna array for small-sized BS. The degradation is due to the high spatial correlation of channels \cite{Rusek_SPMag_12, Lim_ICC14}. A good solution for installing a compact antenna array could involve a collocated dual-polarization antenna system. One dual-polarization antenna could then play, equivalently, the roles of two dipole/patch-type antennas as long as there is high cross-polarization discrimination (XPD). Here, XPD is the ratio of the copolarization received power and the cross-polarization received power.

To establish a dual-polarized MIMO system, we should investigate channel modeling, which is critical for performance evaluation~\cite{DP_WCM}. In prior work \cite{DPchannel, DPchannel2}, the authors focused on the channel from a dual-polarization antenna to a dual-polarization antenna. 
Meanwhile, it is also important, in practice, to study channels from a dual-polarization antenna to dipole/patch-type antennas because a typical MS is likely to be equipped with dipole/patch-type antennas. 

It is necessary to have, as noted above, a high XPD to set up dual-polarized MIMO systems. Researchers have developed dual-polarization antennas that have a high XPD in the main direction of radiation \cite{DP2, DP3}. It is also important, on the other hand, to consider the average XPD in all directions in indoor environments, for signals go through a wall-induced reflection in all directions \cite{ManufDP}. While the conventional approach to designing a dual-polarization antenna with high XPD provides better MIMO performance, it is a costly and complex process to make this antenna at a small size. From this perspective, it remains an open question as to how much XPD is needed to sustain MIMO performance at reasonable cost and level of complexity.

In this letter, we provide a correlated channel model for a dual-polarization antenna to omnidirectional antennas in indoor small-cell MIMO systems. This channel model reflects the relationship of XPD in the direction of angle-of-departure (AoD) of a dual-polarization antenna and correlation of MIMO channel. We evaluate the dual-polarization antenna-based MIMO channel model by using a three-dimensional (3D) ray-tracing simulator that can exploit physically specific behaviors of the polarized channel~\cite{ComMag_METIS,Lim_Mapbased}. We also investigate the equivalent distance between adjacent antennas according to the XPD. We provide insight into the XPD design aspect for a dual-polarization antenna at a small size and at a reasonable cost and level of complexity. To the best of our knowledge, this letter is the first work to explain and validate the relationship between an XPD and spatial correlation by utilizing the presented channel model and a 3D ray-tracing tool. In 3GPP, this problem has been an open issue.

\section{System Model}
\begin{figure}[!t]
\centering{\includegraphics[width=0.75\columnwidth]{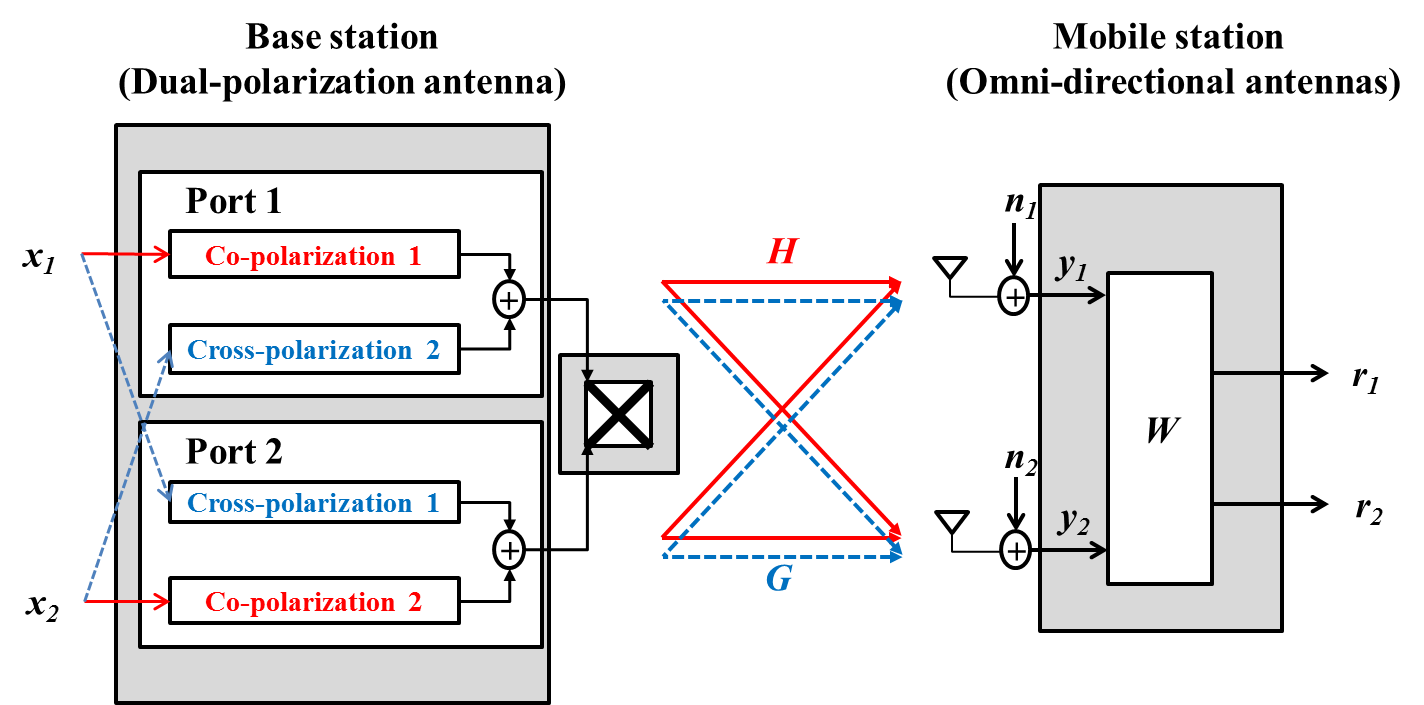}}
\caption{Block diagram of the dual-polarization antenna-based MIMO system.}
\label{systemmodel}
\end{figure}
Fig.~\ref{systemmodel} illustrates a block diagram of a dual-polarization antenna-based MIMO system. We consider a single-user MIMO with one collocated dual-polarization antenna that serves one MS equipped with two omnidirectional antennas. This model is considered a 2$\times$2 MIMO system, since horizontally and vertically polarized waves are respectively radiated from Port-1 and Port-2 of the dual-polarization antenna, the waves go through the channel almost independently due to zero cross-correlation between the orthogonally polarized waves\cite{DP_WCM, ManufDP,3DDPchannel}. Let $h_{rt}$ denote the channel coefficient between the $t^\text{th}$ port at the BS and the $r^\text{th}$ antenna at the MS, which is an independent and identically distributed (i.i.d.) complex Gaussian random variable with zero mean and unit variance. From the channel model with a polarization transfer matrix that consists of random phases for different polarization combinations in \cite{3GPP_3DSCM, 3DDPchannel}, the coefficient of the effective channel ($\pmb{H}_{\text{eff}}$) is given by
\begin{equation}
h_{rt}^\text{eff}=\begin{bmatrix}
\sqrt{L_{r,t}^{-1}}e^{j\Phi^{rH}}\\\sqrt{L_{r,t^{\prime}}^{-1}}e^{j\Phi^{rV}}
\end{bmatrix}^T
\begin{bmatrix}
\sqrt{G_{\phi,\alpha_t}}\\ \sqrt{G_{\phi,\beta_t}}
\end{bmatrix}
=\sqrt{\alpha_{rt}}h_{rt}+\sqrt{\beta_{rt^{\prime}}}h_{rt^{\prime}}\nonumber,
\end{equation}
where $t\neq t^\prime$.\footnote{The conventional dual-polarization channel models in \cite{DP_WCM, DPchannel, DPchannel2} can be obtained by letting $G_{\phi,{\beta_t}}=0$. Since they did not consider the propagation of cross-polarized waves, they applied statistical XPD values to the off-diagonal elements in the polarization transfer matrix.} The notations $\Phi^{rH}$ and $\Phi^{rV}$ are random initial phases, the distributions of which are uniform within ($-\pi,  +\pi$), for the horizontal and vertical polarization combinations at the $r^\text{th}$ MS antenna. In addition, 
$\alpha_{rt}$ and $\beta_{rt}$ are the propagation gains from copolarization and cross-polarization, and $G_{\phi,{\alpha_t}}$ and $G_{\phi,{\beta_t}}$ are, respectively, the antenna gain of copolarization and cross-polarization in the direction of AoD, which is denoted as $\phi$. Finally, $L_{r,t}$ is the~path loss. Thus, the effective channel can be decomposed into the channels from copolarization and cross-polarization, which is given by $\pmb{H}_\text{eff}=\pmb{H}+\pmb{G}$, where
\begin{equation}
\pmb{H}=\begin{bmatrix}
\sqrt{\alpha_{11}}h_{11}&\sqrt{\alpha_{12}}h_{12}\\
\sqrt{\alpha_{21}}h_{21}&\sqrt{\alpha_{22}}h_{22}
\end{bmatrix},\quad\pmb{G}=\begin{bmatrix}
\sqrt{\beta_{12}}h_{12}&\sqrt{\beta_{11}}h_{11}\\
\sqrt{\beta_{22}}h_{22}&\sqrt{\beta_{21}}h_{21}
\end{bmatrix}.\nonumber
\end{equation}

We assume that the large-scale parameters from the $t^\text{th}$ port at the BS are the same, regardless of the antenna index at the MS (i.e., $\alpha_{rt}=\alpha_{t}$, $\beta_{rt}=\beta_{t}$ and $L_{r,t}=L_{t}=L$). This assumption is valid when the MS is a small device~\cite{ComMag_METIS}. The XPD at the MS with the corresponding AoD from the $t^\text{th}$ port of a dual-polarization antenna is then expressed by
\begin{equation}
\chi_{\phi,t}=\frac{\mathbb{E}_{h}[|\sqrt{\alpha_t}h_{rt}|^2]}{\mathbb{E}_{h}[|\sqrt{\beta_{t^\prime}}h_{rt^\prime}|^2]}=\frac{\alpha_t}{\beta_{t^\prime}}=\frac{G_{\phi,\alpha_t}L_{rt^\prime}}{G_{\phi,\beta_t}L_{rt}}=\frac{G_{\phi,\alpha_t}}{G_{\phi,\beta_t}}.
\label{eqXPD}
\end{equation}
%

\section{Correlated Channel Model for a Dual-Polarization Antenna}

From the effective dual-polarization channel, we recognize that despite of the orthogonal polarization, $\pmb{G}$ affects the correlation between its components in the real propagation channel. From~(\ref{eqXPD}), while the formula of XPD is a function of the average received powers ($\alpha_t$ and $\beta_t$) in a certain AoD, XPD can be calculated from the measured radiation pattern in an anechoic chamber. The spatial correlation between adjacent BS antennas depends on the antenna configuration and the AoD in the conventional spatial MIMO channel model. In contrast, since two ports are collocated in the dual-polarization antenna, to investigate such correlation, we consider XPD parameters rather than adjacent antenna spacing.
\begin{figure}[!t]
	\centering{\includegraphics[width=75mm]{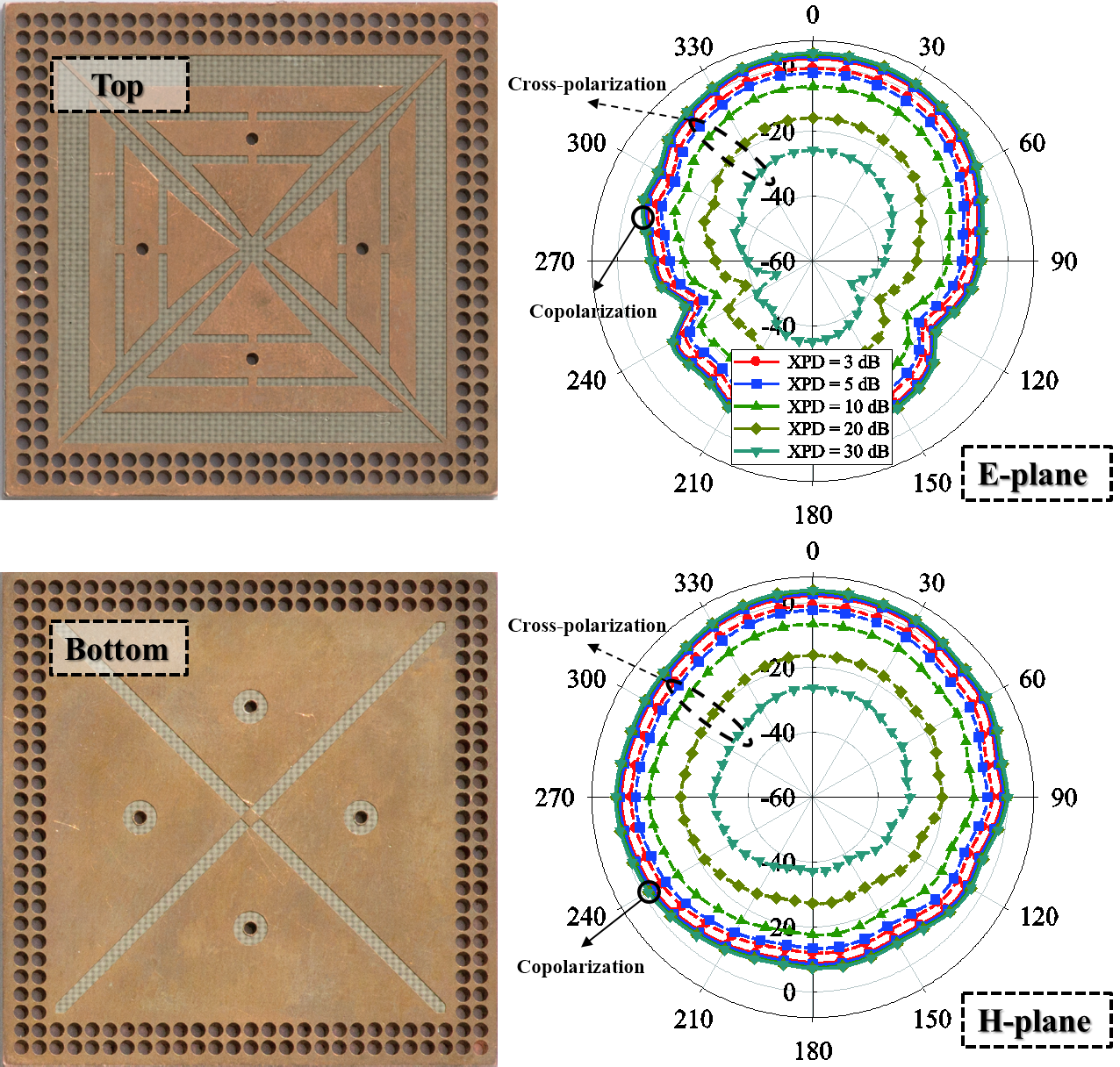}}
	\caption{On the left are pictures of the fabricated dual polarization antenna (up: top view, down: bottom view). On the right are the radiation patterns of different XPD cases (up: E-plane, down: H-plane, solid line (outside): copolarization, dashed line (inside): cross-polarization). The shapes of patterns are measured from the fabricated dual-polarization antenna. Meanwhile, the gains of patterns of each XPD are estimated under the same power constraint.}
	\label{pattern}
\end{figure}




To construct the correlated channel model for a dual-polarization antenna, we first assume that there is a negligible spatial correlation at the MS. This assumption ensures that the effective channel consists of only transmit side effects. From~(\ref{eqXPD}), the effective channel is decomposed into
\begin{equation}\label{Heff}
\pmb{H}_\text{eff}=\begin{bmatrix}
\sqrt{\alpha_{1}}h_{11}&\sqrt{\alpha_{2}}h_{12}\\
\sqrt{\alpha_{1}}h_{21}&\sqrt{\alpha_{2}}h_{22}
\end{bmatrix}\begin{bmatrix}
1&\frac{1}{\sqrt{\chi_{\phi,1}}}\\
\frac{1}{\sqrt{\chi_{\phi,2}}}&1
\end{bmatrix}.
\end{equation}
From (\ref{Heff}), the effective channel can be represented as the Kronecker model (i.e., $\pmb{H}_\text{eff}=\pmb{H}\pmb{R}^{1/2}_{\text{TX}}$), a model that is commonly used to analyze the correlation of channel components~\cite{3DDPchannel,3GPP_3DSCM}. Thus, the effective correlation matrix at the transmitter is expressed by
\begin{equation}
\pmb{R}_{\text{TX(dualpole)}}
\approx\begin{bmatrix}
1&\frac{2}{\sqrt{\chi_{\phi,1}}}\\
\frac{2}{\sqrt{\chi_{\phi,2}}}&1
\end{bmatrix},
\label{eqRTX}
\end{equation}
with an approximation, $\frac{\sqrt{\chi_{\phi,1}\chi_{\phi,2}}+1}{\sqrt{\chi_{\phi,1}\chi_{\phi,2}}}\approx 1$, in the high XPD regime.\footnote{This approximation also holds in the typical range where the average XPDs are~{7-9~dB}~\cite{3GPP_3DSCM}.} Equation (\ref{eqRTX}) implies that the correlation coefficient between the ports can be represented as $\rho_{\textrm{dualpole}}=2/\sqrt{\chi}$ when the XPDs of both ports in the same direction are the same. The effective correlation matrix implies that XPD affects the correlation of channels, which affects the throughput of MIMO. Consider, for example, the worst case of XPD (=~0~dB). Since that yields a rank-1 MIMO channel, spatial multiplexing is not available.

\section{3D Ray-tracing-based System-Level Simulation}

\begin{figure}[!t]
\centering{\includegraphics[width=0.6\columnwidth]{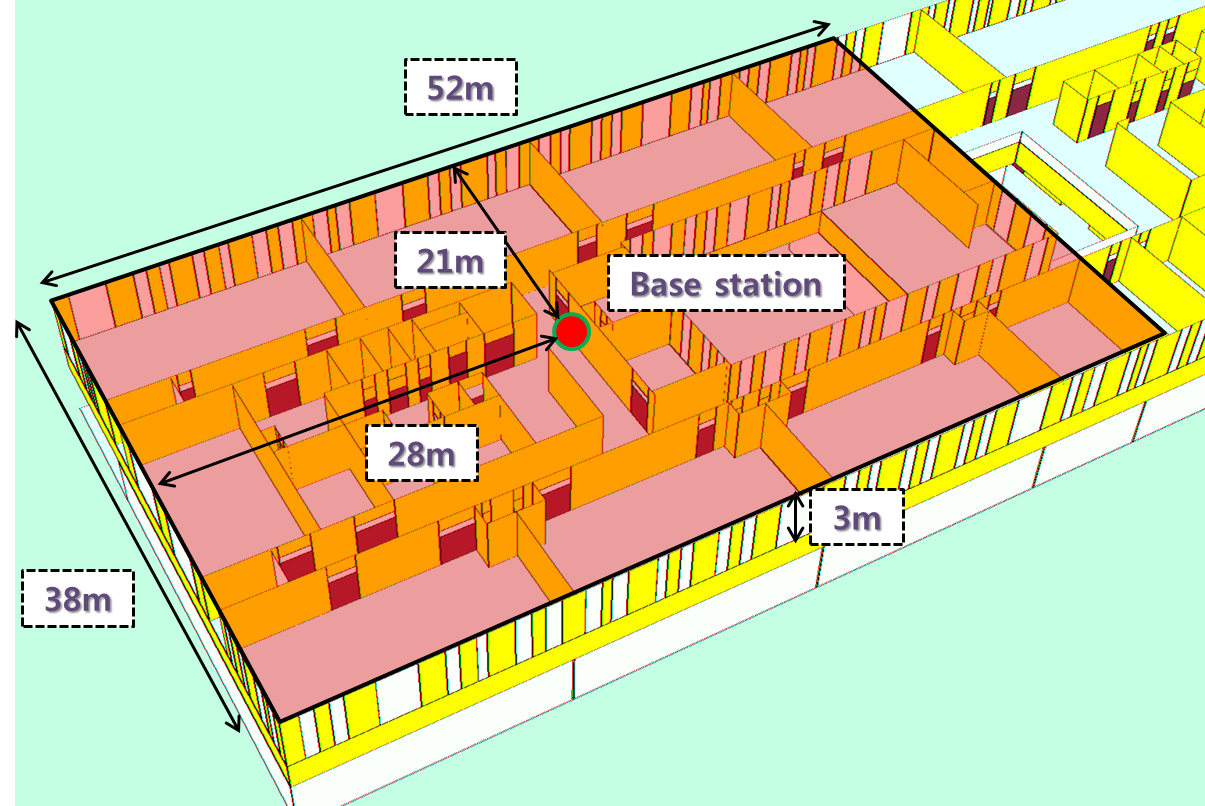}}
\caption{Perspective view of the test site for 3D ray-tracing simulations.}
\label{raytracing}
\end{figure}


We develop a system-level simulator based on a 3D ray-tracing tool (WiSE, Wireless System Engineering developed by Bell Labs)~\cite{WiSE_VTC98}. Fig.~\ref{raytracing} shows the building database of the test site from the perspective view of the second floor of the Veritas C Hall at Yonsei University~\cite{Lim_Mapbased}. The digital map includes concrete walls and floors, metallic doors, glass windows, and sheetrock ceilings. The BS is located immediately under a 3~m-high ceiling in a hallway with one dual-polarization antenna facing the rooms to the southeast. The system parameters are set to those for the Long Term Evolution (LTE) systems; the FFT size = 1024 (8.4~MHz effective bandwidth), the system overhead = 25.22~percent, and the maximum modulation order = 64~QAM (5/6 code rates) \cite{Baker_LTE12}. The propagation gains ($\alpha_{ij}$ and $\beta_{ij}$) are measured by using the 3D ray-tracing tool based on the manufactured antenna pattern~\cite{ManufDP} in Fig.~\ref{pattern}. We assume that, for notational convenience and for investigating the effect of different XPDs on MIMO systems without loss of generality, the shapes of the radiation patterns of both copolarization and cross-polarization are the same.\footnote{This assumption holds for a high average XPD in all directions, which is desired for a MIMO system in an indoor environment~\cite{ManufDP}, and the extension for any pattern is straightforward.} We also measure the propagation gain ($\alpha_\text{omni}$) from the BS with two omnidirectional antennas (for the conventional antenna configuration). 
From these measurements, we construct channel matrices $\pmb{H}$ and $\pmb{G}$. 
Finally, we evaluate the average throughput in four effective channel models: i) the whole measured channel from the dual-polarization antenna ($\pmb{H}+\pmb{G}$); ii) the analytical channel with the effective correlation matrix ($\pmb{H}\pmb{R}_\text{TX(dualpole)}^{1/2}$); iii) the whole measured channel with the spatial correlation matrix of two omnidirectional antennas ($\pmb{H}_\text{omni}\pmb{R}_\text{TX(omni)}^{1/2}$); iv) the analytical channel with the effective correlation matrix assuming the omnidirectional pattern ($\pmb{H}_\text{omni}\pmb{R}_\text{TX(dualpole)}^{1/2}$), where $\pmb{H}_\text{omni}$ is the channel matrix whose elements are denoted by $\sqrt{\alpha_\text{omni}}h_{rt}$.

We assume a zero-forcing receiver described as $\pmb{W}^T = \left(\pmb{H}_\text{eff}\right)^{-1}$ (i.e., open-loop spatial multiplexing assuming perfect channel estimation at the MS). Thus, the signal-to-interference-plus-noise-ratio (SINR) of the $i^\text{th}$ received signal is $\text{SINR}_i=\frac{1}{||\pmb{w}_i||^2\text{p}_\text{n}}$, where $\pmb{w}_i$ and $\text{p}_n$ are the $i^\text{th}$ column vector of $\pmb{W}$ and the background noise (-174~dBm/Hz), respectively. The average throughput is calculated from the results of 1,000 simulations for 802 differently located users of which the distribution is uniform within 1~m above the floor in the shaded region of the figure, assuming a single-user MIMO system (the BS serves only one MS at a time).
\subsection{Comparison of the Measured and Proposed Channels}
Fig. \ref{result} shows the cumulative distribution function (CDF) of the average throughput extracted from our system-level simulator by the effective channel models i) and ii). It shows that the higher the XPD of the dual-polarization antenna, the higher the throughput of the MIMO channel it provides. Besides, the average throughput of the analytical channel well approaches that of the whole measured channel. There are small gaps between them because the 3D ray-tracing simulator measures accurate polarized-propagation.

\subsection{Relationship between XPD and Antenna Spacing}
In this section, we analyze the relationship between the antenna spacing of two omnidirectional antennas and the XPD of one dual-polarization antenna, the throughputs of which are the same.
Such a relationship in the manner of the correlation comparison has not been investigated, while prior work has focused on the relationship between the cross-correlation of each polarization and the slant angle~\cite{DP_WCM,3DDPchannel}. 

To obtain the average throughputs, we use $\pmb{H}_\text{omni}\pmb{R}_\text{TX(omni)}^{1/2}$ where two types of AoD distribution, isotropic and Laplacian, are considered for calculating $\pmb{R}_\text{TX(omni)}$. Isotropic distribution is usually used for the rich-scattering environments while Laplacian distribution reflects the real measurements of AoDs of corresponding MSs through the 3D ray-tracing simulation. From the given AoD distribution, we get the spatial correlation coefficient of the $t^\text{th}$ row of ${\pmb{R}_\text{TX(omni)}}$ as
$\rho_{\text{omni},t} = \mathbb{E}_{\phi}[e^{-jkd\sin\phi}]$,
where $k$ is the wavenumber and $d$ is the distance between adjacent antennas. 
Then, for each MS, the effective channel with the measured spatial correlation matrix of two omnidirectional antennas is derived as
\begin{equation}
\heffl=\sqrt{\frac{p_\ell}{\sum_{\ell=1}^{L}p_\ell}}\sum\limits_{\ell=1}^{L}{{\pmb{H}_{\text{omni},\ell}}({\pmb{R}_{\text{TX(omni)},\ell}^{1/2}})^T},
\label{heff_iso}
\end{equation}
where $p_\ell$ is the power-delay profile for the $\ell^\text{th}$ tap and $L$ is the total number of the channel taps for an MS, which are measured from the 3D ray-tracing simulation.

\begin{figure}[!t]
\centering{\includegraphics[width=0.73\columnwidth]{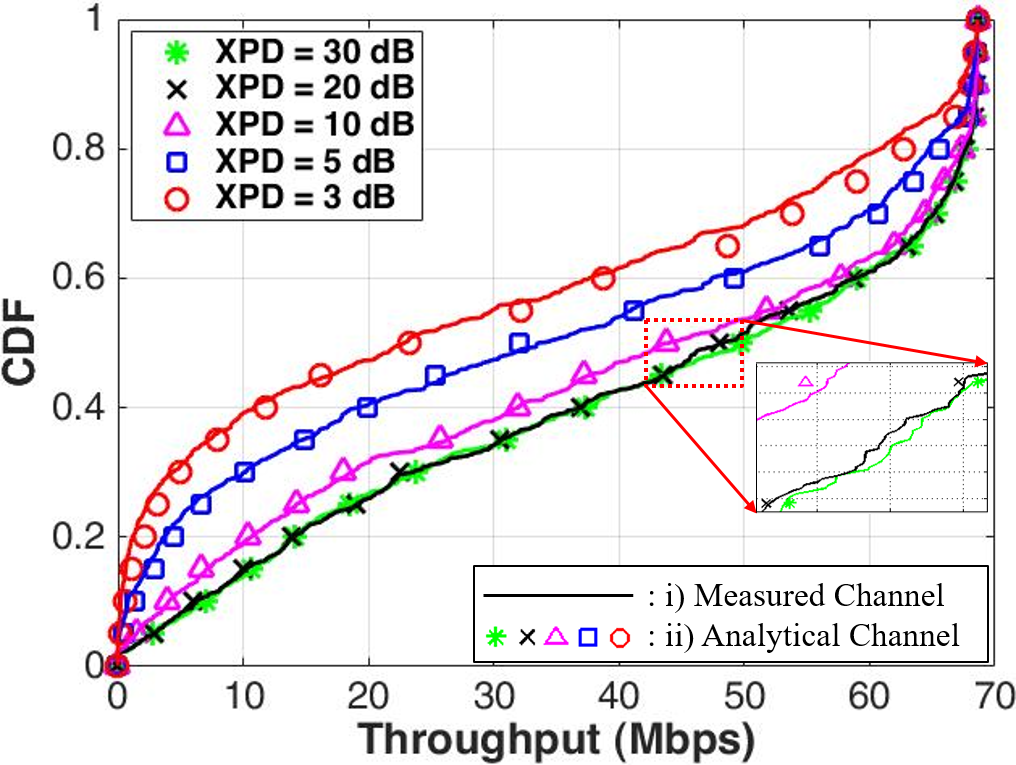}}
\caption{CDF of the ergodic throughput of 3D ray-tracing simulation for the effective channel models i) and ii).}
\label{result}
\end{figure}

\begin{figure}[!t]
\centering{\includegraphics[width=0.73\columnwidth]{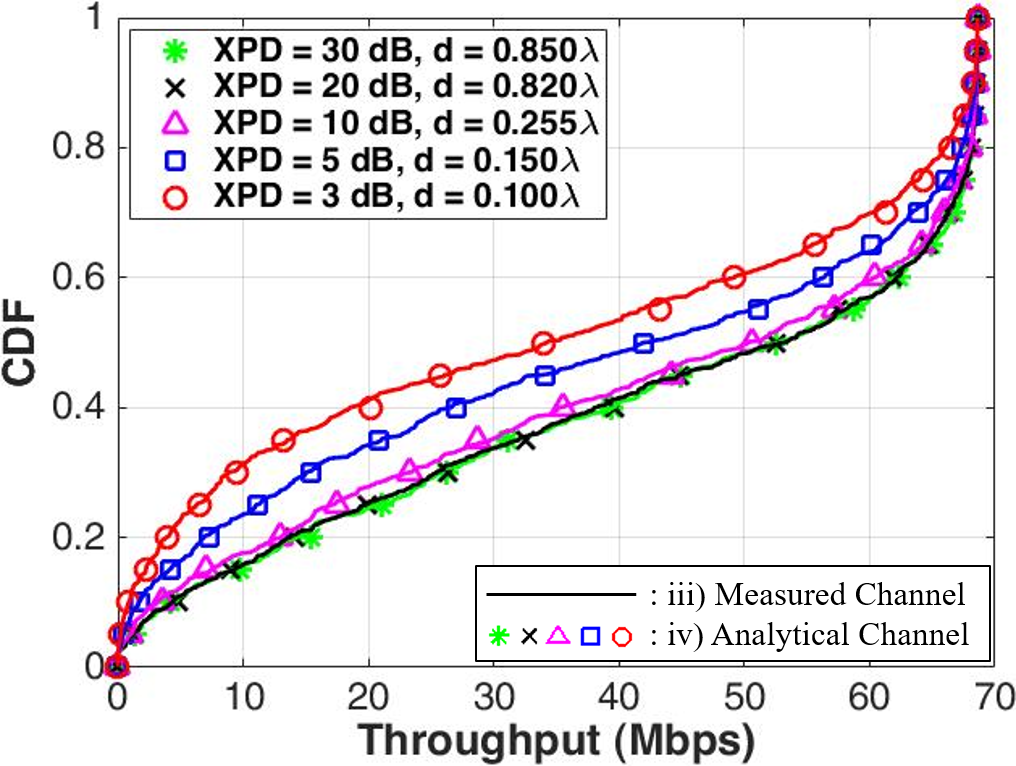}}
\caption{CDF of the ergodic throughput of 3D ray-tracing simulation for the effective channel models, iii) with Laplacian AoD distribution and iv). The antenna separation ($\dlap$) corresponding to the XPD are given in Table I.}
\label{result_SCM}
\end{figure}

Fig.~\ref{result_SCM} shows a comparison of the CDF of the average throughput calculated from effective channel models, iii) with Laplacian AoD distribution and iv). It shows that there is a relationship between the XPD and the spatial correlation and this affects the throughput.
In Table~\ref{Table1}, we show an example (for our simulation set up) that compares the derived equivalent separation between adjacent omnidirectional antennas (i.e., $\diso, \dlap$) and the effective correlation coefficient, $\rho_{\textrm{dualpole}}$, for different XPD values. In can be seen that as $\rho_{\textrm{dualpole}}$ decreases with the increasing XPD, $\diso$ and $\dlap$ increase.
Note that the gaps of the equivalent separation between XPD = $20, 30$~dB in both the cases of $\diso$ and $\dlap$ are much smaller than the others, which supports the result in Fig.~\ref{result}. 

\subsection{Antenna Design Aspect}
From the relationship between the XPD and the spatial correlation, we can gain two valuable insights into antenna design. First, our research shows a negligible difference in throughput between the cases of XPD = $20$~dB and $30$~dB.
This result implies that it may be inefficient to make a very high XPD-dual-polarization antenna for MIMO, given the high complexity and cost of such an endeavor for small BSs or small devices.
Second, using the equivalent separation, we can simply design various uniform arrays, which are organized by dual-polarization antennas. In this section, we use as an example the equivalent separation derived in Table~\ref{Table1}. If we make a uniform planner array with dual-polarization antennas with $20$~dB of XPD for the compact antenna array, the distance between the adjacent dual-polarization antennas will be $0.820 \lambda$.
Therefore, the presented effective correlation matrix could be a meaningful tool for both evaluating and designing dual-polarization antennas for a compact antenna array in indoor small-cell MIMO systems.

\section{Conclusion}

\renewcommand{\arraystretch}{1.3}
\begin{table}[!t]
	\caption{Equivalent spatial separation between adjacent antennas compared with XPD}
	\vspace{-0.6cm}
	\begin{center}
		\resizebox{1\columnwidth}{!}{
			\begin{tabular}{|c||c|c|c|c|c|}
				\hline
				\rowcolor{lightgray}\textbf{XPD (dB)} & \textbf{3} & \textbf{5} & \textbf{10} & \textbf{20} & \textbf{30} \\ 
				\hline
				$\rho_{\textrm{dualpole}}$ (Eq. (3)) & 0.9432 & 0.8545 & 0.5750 & 0.1980 & 0.0632\\
				\hline
				$\dlap$ (Fig.  5) & 0.100$\lambda$ & 0.150$\lambda$ & 0.255$\lambda$ & 0.820$\lambda$ & 0.850$\lambda$ \\
				\hline
				$\diso$ & 0.076$\lambda$ & 0.124$\lambda$ & 0.220$\lambda$ & 0.326$\lambda$ & 0.364$\lambda$\\
				\hline
		\end{tabular}
	}
	\end{center}{\label{Table1}}
\vspace{-0.4cm}
\end{table}

In this letter, we have presented a correlated channel model for a dual-polarization antenna to omnidirectional antennas for indoor small-cell MIMO systems. The presented effective correlation matrix of a dual-polarization channel reflects the relationship between the XPD and the spatial correlation. From the analysis and the results through extensive 3D ray-tracing-based simulations, we have confirmed that the presented effective correlation matrix could be a potential tool for both evaluating and designing dual-polarization antennas. In future work, we will extend our research to various environments and systems. We will also analyze the cost and the complexity of manufacturing a high XPD-dual-polarization antenna.

\bibliographystyle{IEEEtran}
\bibliography{XPD_reference_etal}

\end{document}